\documentclass[a4paper,11pt]{article}
\usepackage{jinstpub}
\title{An eight-channel RFSoC waveform digitizer with \mbox{AC- and DC-coupled} front ends and transparent dual-DDR event memory}
\author[a,b,1]{Tao Xue\note{Corresponding author.}}
\affiliation[a]{Key Laboratory of Particle \& Radiation Imaging (Tsinghua University), Ministry of Education,\\
Qinghuayuan Street No.~1, Beijing, China}
\affiliation[b]{Department of Engineering Physics, Tsinghua University,\\
Qinghuayuan Street No.~1, Beijing, China}
\emailAdd{xuetaothu@tsinghua.edu.cn}

\abstract{We present Transparent Dual-DDR Event Memory (TDEM) for
the triggered readout of an eight-channel radio-frequency system-on-chip
(RFSoC) digitizer. With either an AC- or a DC-coupled front end, each
accepted trigger records 4096 samples per channel at 5~GSa/s.
Two pulses 100~ns apart produced separate, overlapping records.
On one DC-coupled input, interleaving calibration reduced an eight-sample
periodic residual by a factor of 29 at $+100$~mV, but not at zero input.
After a power cycle, the digitizer restored the coefficients for all eight
DC-coupled inputs without a training tone. TDEM assigns complete events to
64- and 32-bit DDR4 memory interfaces in a 2:1 ratio. With this allocation
alone, host page requests still determined which memory served each read.
In two runs of approximately 6~s at 160~kHz, the peak receiver backlog fell from
20\,140 events with the earlier readout to 1457 with event-level prefetch.
Prefetch raised the sustained rate of the DC-coupled system from 150 to
170~kHz under repeated host readout pauses. With the same receiver design
and a revised host buffer allocation, the AC-coupled system passed three
30~min tests at 175~kHz, each with 48 readout pauses and no rejected
triggers. Without imposed pauses or background load, the host verified all
323\,452\,631 events in 30~min at 179.7~kHz. At 180~kHz, triggers were
first rejected at the digitizer, consistent with a modelled transmitter
capacity of 179.730~kHz for this event format.}

\keywords{Front-end electronics for detector readout; Data acquisition circuits; Data acquisition concepts}
\begin{document}
\maketitle

\section{Introduction}
Coincident pulses on several detector channels can carry information that
is lost when only a timestamp or a pulse integral is retained. A waveform
also records the baseline before a pulse and may reveal a second pulse
within the same acquisition window. Switched-capacitor samplers have long
provided this capability in particle instrumentation~\cite{Ritt2010DRS},
and waveform-based timing studies illustrate the importance of distinguishing
measurements with electrical test pulses from those with a
detector~\cite{Breton2016SAMPIC}. Here we measure the rate at which a
digitizer delivers complete eight-channel waveforms to a host and identify
the first stage that rejects triggers as the input rate is increased.

Radio-frequency system-on-chip (RFSoC) devices integrate data converters and
programmable logic on one chip.
Axani et al.\ demonstrated RFSoC pulse acquisition
for scintillator readout and characterized its analogue front end, digital
processing and transfer rate~\cite{Axani2024Pulse}. Using the same class of
device for a different purpose, Du et al.\ combined modular RF front ends,
synchronization and deterministic feedback for accelerator
controls~\cite{Du2025LLRF}. Optical interfaces implemented in
field-programmable gate arrays (FPGAs) can transfer detector data to a
commodity server over PCI Express (PCIe), as demonstrated by
FELIX~\cite{Anderson2016FELIX}. The ATLAS RobinNP receiver uses two
on-board memories, each serving a fixed group of six input links
through its own readout logic~\cite{Borga2023ROS}. Reordering accesses
to improve dynamic random-access memory throughput is also established in FPGA
accelerator design~\cite{Csordas2019Reordering}. Here a single stream of
events must use two unequal memory interfaces without relying on either
the source or the host to schedule accesses to individual memories.

An earlier report introduced the ZYNQBeeRF7 module and measured its
converter response on a prototype~\cite{Wei2026TAUP420}. The present work
uses that module with interchangeable active DC-coupled and passive
AC-coupled front ends. Its principal readout contribution is Transparent
Dual-DDR Event Memory (TDEM): the RDX051 receiver assigns, commits and
fetches whole events across two DDR4 memories, while the host requests
logical pages without specifying a memory interface. We compare three
implementations to distinguish placing events across both memories from
keeping both memories supplied with useful reads. We then separate this
receiver change from a later change to host memory allocation. Calibration
and analogue measurements establish what each retained waveform represents.
We distinguish the known inputs needed to obtain converter corrections from the
conditions needed to restore them before acquisition. All measurements
reported here use electrical bench stimuli, not detector data.

\section{Digitizer and capture architecture}
\label{sec:architecture}
\begin{figure}[tbp]
\centering
\includegraphics[width=\linewidth]{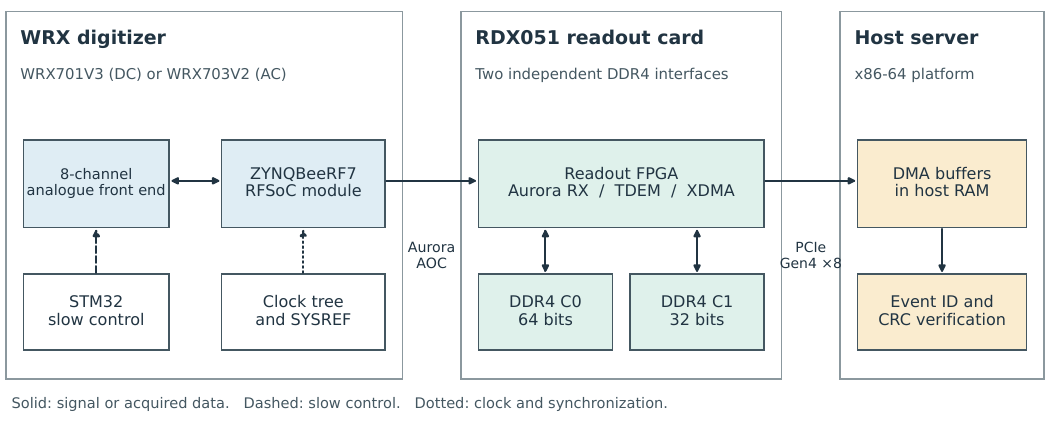}
\caption{Bench system for analogue characterization and triggered
readout. Either WRX carrier accepts the same ZYNQBeeRF7 module. ADC
events traverse one Aurora link over a nominal 100~Gbit/s active optical cable (AOC); DAC test
signals are measured at the carrier, not transmitted through that link.
The control processor is outside the per-event data path.}
\label{fig:hardware}
\end{figure}

Figure~\ref{fig:hardware} shows the common readout. The ZYNQBeeRF7 module
contains an AMD Zynq UltraScale+ XCZU47DR RFSoC. Eight real
analogue-to-digital converter (ADC) channels operate at 5~GSa/s per channel
without decimation. Each 14-bit sample is left-aligned in a signed
16-bit container; four container counts correspond to one converter code.
The eight digital-to-analogue converter (DAC) channels also support
independent real samples at 5~GSa/s without interpolation in the mode
used here. A shared clock tree supplies the sampling clocks and a
system-reference signal (SYSREF) for multi-tile synchronization (MTS).
On start-up the converter tiles are initialized and their MTS status is
checked before capture is enabled~\cite{AMD2025RFDC}. The sample interval
is 200~ps; it is not a measurement of interchannel skew or absolute timing
resolution.

The WRX701V3 DC-coupled carrier uses a fully differential LMH5401
amplifier. A precision DAC80508 and an OPA2192 set the operating
point independently of the detector signal. This path can measure a
static level as well as a pulse. On the WRX703V2 AC-coupled carrier,
RF transformers perform single-ended to differential conversion at the
inputs. Channels 0--3 use TCM2-33WX+ transformers and channels 4--7 use
JY-TCM2-33WX+ transformers; the same grouping is used for the outputs.
Since transformer type and channel position were not interchanged, a
systematic difference between groups would not by itself identify a
component effect.

The capture logic accepts a threshold crossing, an external trigger
or a software trigger. In high-rate operation one accepted trigger
selects a common time window of 4096 samples on each of the eight
channels. Two capture banks maintain the sample history. Once triggered,
a bank completes its window and holds it for readout; the other bank can
accept a second trigger if it has sufficient history. A bank becomes
eligible again only after its final word has been accepted into the
local transmit path and the required history has refilled. This release
does not wait for the host. If neither bank is eligible, no third record
is allocated; the source counts the busy rejection explicitly.

The event size depends on where it is counted (table~\ref{tab:bytes}).
The 64~KiB of samples acquire framing overhead at the Aurora
64B/66B user interface~\cite{AMD2024Aurora}; the receiver uses a
different, page-aligned representation for host reads. For example,
at 179.7~kHz the logical write and read budgets sum to 24.313~GB/s.
This is an event-size budget, not a measurement of either controller's
bus utilization. Data rates in GB/s and Gbit/s are decimal; KiB, MiB
and GiB denote binary multiples of the byte.

\begin{table}[!htbp]
\centering
\caption{Bytes per accepted event at four interfaces. The rows describe
different representations; only the last two should be added for a
logical memory read-plus-write budget.}
\label{tab:bytes}
\small
\begin{tabular}{l r l}
\hline
Representation & Bytes & Contents or unit of transfer\\
\hline
ADC waveforms & 65\,536 & 8 channels $\times$ 4096 samples $\times$ 2 bytes\\
Aurora user stream & 68\,096 & 2128 user words of 256 bits, including framing\\
Receiver memory write & 65\,664 & Sample payload and receiver metadata\\
Logical host read & 69\,632 & 1 descriptor page and 16 payload pages\\
\hline
\end{tabular}
\end{table}

\section{Transparent Dual-DDR Event Memory}
\label{sec:tdem}
\subsection{From a single memory interface to proportional event placement}
\begin{figure}[htbp]
\centering
\includegraphics[width=\linewidth]{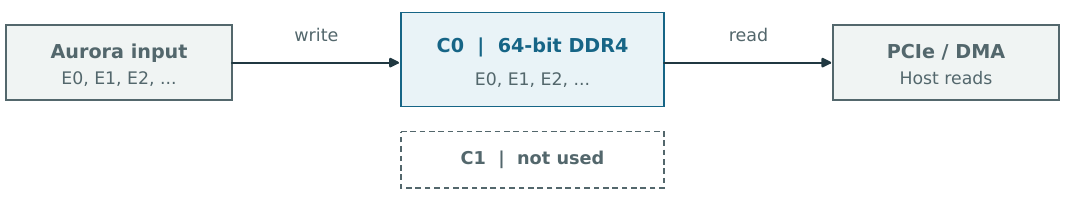}
\caption{Before TDEM, three successive events $E_0,E_1,E_2$ used
the same memory interface in both directions. C1 was present on the
card but did not service a waveform read or write. Arrows show data flow.}
\label{fig:single}
\end{figure}

Each accepted trigger produces a complete eight-channel record. While
one event is arriving, earlier events must be read from the receiver.
In the first implementation both transfers used the 64-bit DDR4
interface (C0); the independent 32-bit interface (C1) carried no
event traffic (figure~\ref{fig:single}). In this section a \emph{bank}
means the memory on one of these two DDR4 interfaces, not a bank inside
a DDR4 device or a capture bank in the digitizer. At 120~kHz the
receiver's input first-in, first-out buffer (FIFO) overflowed when
direct memory access (DMA) reads overlapped
writes, with either two or four readers. This was a limit of that
configuration, not a measurement of the maximum rate of a DDR4 device.

TDEM instead assigns successive admitted events to C0, C0, C1 and
repeats. Each event stays in one bank. The index is the receiver's
admission count, not the trigger number, which may have gaps. Of the
61\,440 event slots, 40\,960 are in C0 (4~GiB) and 20\,480 in C1
(2~GiB). The two writes can finish out of order; only a contiguous
sequence of completed events becomes readable. In the first dual-bank
implementation the host library still translated logical events into
physical bank addresses. On a Xeon server, this design passed 130~kHz
for 60~s and 140~kHz in three 30~min pause tests. Thus 130~kHz was
already a \emph{dual-bank} measurement, not a single-bank control.

On the readout host used below, the initial qualified rate was 135~kHz.
Buffering complete bursts at the memory-to-DMA interface, using 32~KiB
packet FIFOs and a pipeline stage on the narrower path, raised it to
150~kHz. At a fixed 150~kHz input, mixed-traffic read rates were
10.56, 11.81 and 10.59~GB/s for the original, modified and restored
receivers, respectively, with the same test DMA configuration. This
comparison measures the combined change, not the effect of one FIFO alone.

\subsection{Page requests can unbalance reads despite proportional placement}
\begin{figure}[htbp]
\centering
\includegraphics[width=\linewidth]{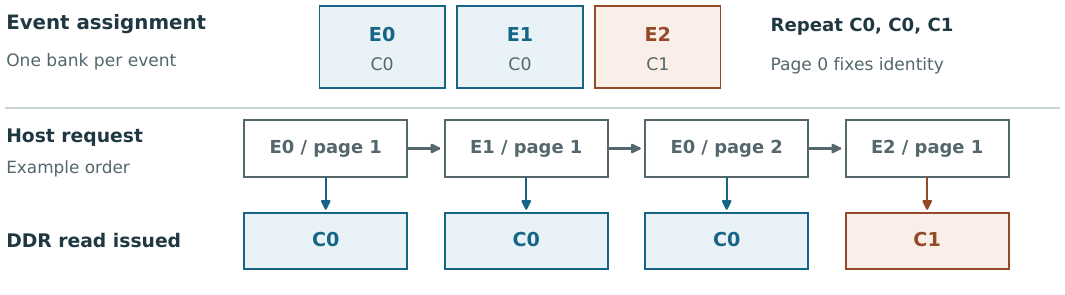}
\caption{Dual-bank allocation without prefetch. Top: strict placement of
three events. Bottom: an illustrative order of host requests after page~0
has been claimed for each. The resulting bank requests follow the host
even though placement is exactly 2:1. The sequence is schematic, not a
recorded trace.}
\label{fig:demand}
\end{figure}

Consider three committed events: $E_0$ and $E_1$ in C0, $E_2$ in C1.
The host sees each as 17 pages of 4~KiB: a descriptor followed by
16 payload pages. An intermediate receiver assigned
events implicitly on the first page request; no reader selected a bank.
But if two readers next requested page~1 of $E_0$ and $E_1$, both
requests still went to C0. Until a reader requested another page of
$E_2$, C1 could do no further read work on it
(figure~\ref{fig:demand}). In this implementation the ring filled after
23.84~s at 150~kHz: of 3\,578\,431 events sent, 3\,577\,553 were
retained and 878 were dropped at the receiver. This is an observed
failure with correct event assignment, not a simulated example.

Traces from the earlier, bank-aware host library show why this distinction
matters. A batch of 3840 events contained exactly 2560 in C0 and 1280
in C1. At 160~kHz, separate snapshots of the DMA read-address interface
recorded 512:170 requests in a 218~$\mu$s window early in a batch and
344:344 in a 281~$\mu$s window later. Neither snapshot is the ratio of
completed events or a measurement of internal memory utilization.

\subsection{Prefetch by event, return by page}
\begin{figure}[htbp]
\centering
\includegraphics[width=\linewidth]{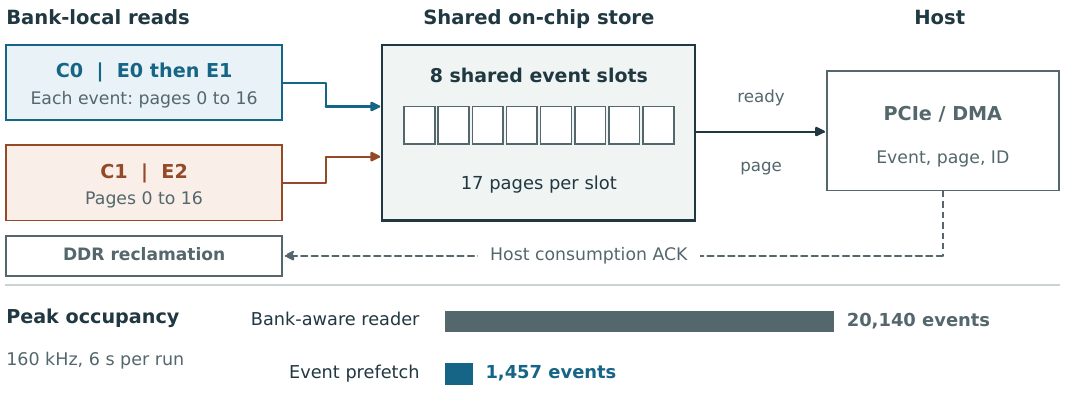}
\caption{Final TDEM. Each memory can now read ahead within an event
independently of host page order. Solid arrows carry page data; the
dashed path represents host acknowledgement of consumption. The slot
holding the original event and the on-chip copy have different lifetimes.
The bars compare peak receiver occupancy in two runs of approximately
6~s at 160~kHz: the earlier bank-aware readout and the final prefetch
design. This comparison is distinct from the 150~kHz test of allocation
without prefetch.}
\label{fig:prefetch}
\end{figure}

The final receiver plans reads for \emph{committed, authorized} events,
not just for pages already requested. Each bank issues the 17 pages of
its current event before selecting its next event; up to eight pages per
bank may be in flight. If only one bank has work it need not wait for
the other to satisfy a per-cycle quota. Eight shared on-chip slots of
68~KiB hold the prefetched data (figure~\ref{fig:prefetch}). The host
can receive a ready page without waiting for the other 16. The response
must match its requested event, page and transaction ID. Prefetch does
not release the original DDR slot: reclamation advances only over
consecutive events acknowledged by the host.

At 160~kHz, two approximately 6~s tests on the same DC-coupled system
gave peak queues of 20\,140 events with the burst-buffered, bank-aware
readout and 1457 with event prefetch. Eight new slots cannot account for
a difference of 18\,683 events by storage alone. The writer, DDR
controllers, memory-side FIFOs and clocks were unchanged. The bank-aware
comparison used a different host interface; the intermediate receiver
that dropped 878 events used the \emph{same} logical reader as the final
design, but was tested in a separate session. These comparisons support
improved read scheduling without assigning a separate gain to each
implementation detail. The final receiver sustained 170~kHz in the
three pause tests described below.

\section{Calibration and operation}
\label{sec:calibration}
\begin{figure}[htbp]
\centering
\includegraphics[width=\linewidth]{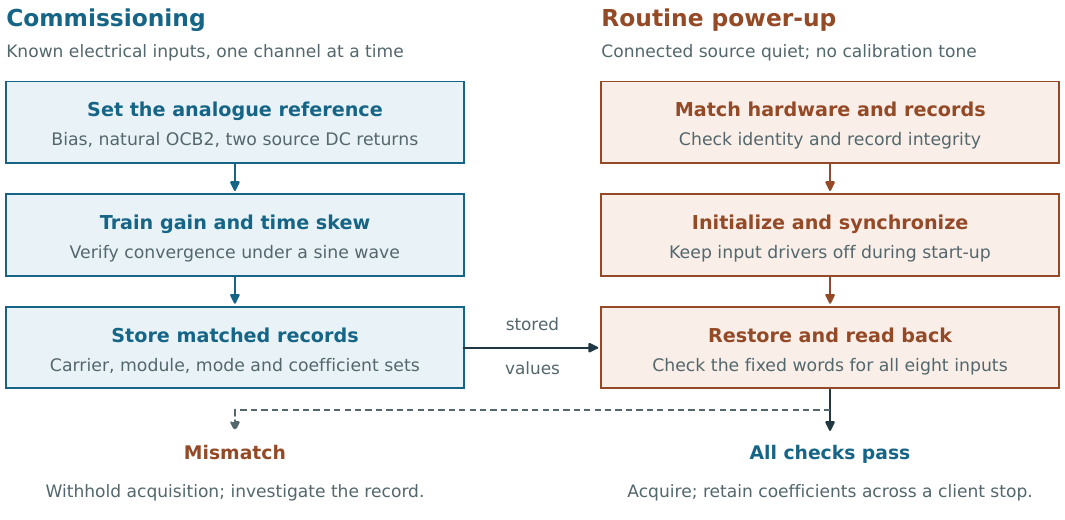}
\caption{Calibration of the DC-coupled digitizer. Left: measurements made
when commissioning a carrier and its RFSoC module, or after a change that
requires recalibration. Right: the sequence performed on power-up. The
horizontal arrow carries stored parameters, not a calibration waveform.
Both a matching configuration and coefficient readback are required to
enable acquisition.}
\label{fig:cal-flow}
\end{figure}

A detector may deliver a steady baseline and only occasional pulses. Neither
condition supplies a known zero or a reliable signal for estimating
interleaved ADC gains. Conversely, a clean spectrum measured after several
seconds of sinusoidal excitation need not describe the state in which the
first detector pulse would have been recorded. We therefore separate two
operations on the DC-coupled digitizer: measuring a set of coefficients with
known inputs, and restoring those coefficients before enabling acquisition
(figure~\ref{fig:cal-flow}). No sine source is required for the second operation.

\subsection{Establishing the reference on the bench}
We first measured the ADC response to small changes of the precision
DAC80508 bias setting. The working point of the LMH5401 input stage was
paired with the second offset-calibration block (OCB2) value obtained by
the converter at that point, leaving adjustment range on both sides.
With this pair fixed, we determined a zero setting for each of two
external DC-return paths. The only connection to the input under test was
a WavePro404HD oscilloscope set to DC coupling and either 1~M$\Omega$
or 50~$\Omega$; no signal was applied. The 700 stored frames cover eight
channels and two settings per channel. Across the 16 settings, the largest
absolute value of a confirmation mean was 16.78 container counts.
Selecting a stored zero accounts for the source's \emph{DC} return path;
it neither changes the RF input impedance nor subtracts a physical offset
carried by a detector.

The RF data converter (RFDC) estimates gain and sampling-time mismatch
from a changing input using its background gain and time-skew calibration
blocks (GCB and TSCB)~\cite{AMD2025RFDC}. A Rohde \& Schwarz SMA100B supplied a
100.317~MHz sine wave to one channel at a time. Its setting was $-12$~dBm;
the digitized tone levels were $-16.59$ to $-16.46$~dB relative to
full scale (dBFS). We fitted a
sine wave, including a constant term, separately to samples with index
$n\bmod 8=j$, using their actual sampling times. If $A_j$ is the fitted
amplitude for class $j$ and $\bar A$ the mean of the eight amplitudes, the
relative gain dispersion is the root mean square (RMS) of $A_j/\bar A-1$.
These are classes of output samples; the analysis requires no assignment
of their indices to physical sub-ADCs.

We allowed the background estimates to converge under the tone and then
froze them. In two successive four-frame blocks the dispersion had to be
at most 0.05\% and the block means had to agree to within 0.03 percentage
points. We saved the two coefficient blocks for that channel and checked
the result after reloading them: all words and override controls had to
match, the dispersion still had to be at most 0.05\%, and the measured
signal-to-noise ratio (SNR) under the same tone could not fall by more
than 0.5~dB. These are acceptance criteria for this system, not converter
specifications. The final check used measurements made after reloading
the coefficients, separate from those used to stop adaptation.

\subsection{Using another stimulus to test the correction}
We then replaced the training tone with successive constant levels from a
DG922 Pro generator. For the comparison before and after restoring the
coefficients, we used the same source connection, analogue settings and
sequence of voltages on input~0. For a constant input, let $\bar x_j$
denote the mean for class $j$ and $\bar x$ the frame mean. We define the
eight-sample periodic residual, in signed 16-bit container counts, as
\begin{equation}
 R_8=\left[\frac{1}{8}\sum_{j=0}^{7}(\bar x_j-\bar x)^2\right]^{1/2}.
 \label{eq:r8}
\end{equation}
We applied eq.~\eqref{eq:r8} to each frame before averaging over the 16 frames at
each input level; pooling samples across frames could cancel a pattern
whose phase changes. At $+100$~mV, $R_8$ fell from 265.22 to 9.11 counts;
at $-100$~mV, from 257.80 to 9.54. In contrast, the initial zero-input
value changed from 6.61 to 10.58 counts (figure~\ref{fig:cal}(a)).

The difference matters for a DC-coupled detector. In a linear model the
class mean at voltage $V$ is $g_j V+o_j$, with gain $g_j$ and offset
$o_j$~\cite{Kurosawa2001Mismatch}. Equalizing the gains reduces a periodic
term that grows with $|V|$; it need not improve the zero-input residual.
We did not subtract a waveform template. Nor does this DC test determine a
timing error, which acts on a signal with nonzero slope. On the AC carrier
we separately exercised input~0 under a training tone. The eight-channel
AC spectral measurements did not use a fixed coefficient table qualified
on all eight inputs by this procedure.

\subsection{DAC output offset}
\begin{figure}[htbp]
\centering
\includegraphics[width=\linewidth]{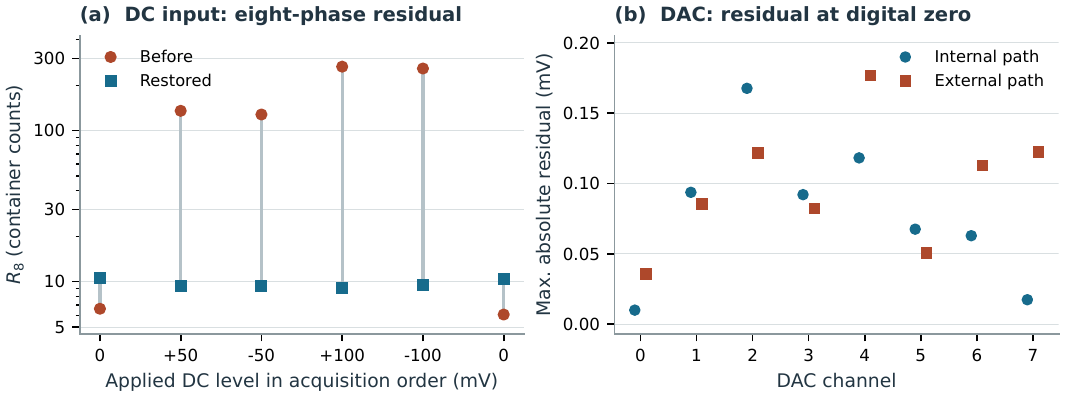}
\caption{Static checks on the DC carrier. (a) Eight-phase RMS residual
on input 0 before and after restoring the learned coefficients; the two
zero-input points bracket the DC sequence (16 frames per point).
(b) Largest absolute value of three independent waveform means at digital
zero for each RF-DAC channel and analogue path, with a 50~$\Omega$ load.}
\label{fig:cal}
\end{figure}

The precision bias DAC used at the ADC input is distinct from the eight
RF-DAC waveform outputs. On the DC carrier, we measured the RF-DAC
quadrature modulation correction (QMC) offset independently for two
selectable analogue paths on every output. With a constant digital zero
and a DC-coupled 50~$\Omega$ oscilloscope input, we measured the mean
output voltage at codes bracketing zero. A local fit $V(q)=a q+b$ gave
$q_0=\mathrm{round}(-b/a)$. Three independent waveforms per path verified
the predicted setting before the original setting was reapplied. Across
178 waveforms in the search and verification sequence, the largest
absolute confirmation mean was 0.177~mV (figure~\ref{fig:cal}(b)). These
path-specific offsets are restored with the static configuration; the
RF-DAC outputs remain muted by default during ADC acquisition. The test
does not calibrate DAC gain, phase or harmonic distortion. A transformer
output cannot transmit the static voltage used for this procedure.

\subsection{Power-up and acquisition without a training source}
\label{sec:cal-operation}
The commissioning record is specific to the carrier, the converter module
and its operating mode. The carrier's nonvolatile ferroelectric memory
holds the static calibration; the digitizer also stores the associated
converter coefficients and their configuration identity. On power-up,
the controller checks the static record and its integrity before enabling
the analogue supplies and clocks. With the input drivers still disabled,
the digitizer executes the converter's normal foreground initialization
and MTS. Successful tile synchronization does not establish that
interleaving gains have converged.

The acquisition service restores the matched analogue settings and their
recorded OCB2 reference. It loads the saved GCB and TSCB words and reads
back all eight channels before enabling capture. A missing or mismatched
record does not cause the system to learn a replacement from the detector.
An ordinary acquisition stop or client disconnect does not release the
fixed coefficients. Freezing alone does not preserve a value across
power loss; once the stored coefficients are reloaded as overrides, the
corresponding blocks no longer adapt to the incoming signal. A converter
reset requires restoration; replacing the
carrier or module, or changing the sampling mode, requires requalification.
Thus the operator need not decide whether a quiet interval is a valid
calibration signal.

This sequence does not replay the entire foreground state. OCB1, the
first offset-calibration block, retains its native behaviour. AMD
describes coefficient override as distinct from pausing adaptation and
discourages manual changes to the offset blocks~\cite{AMD2025RFDC}.
Reapplying the naturally obtained OCB2 value with its analogue working
point is a board-specific choice tested here, not a general prescription.
An attached source is also not necessarily a quiet source: the input
driver's power-down control is not electrical isolation.

\begin{figure}[htbp]
\centering
\includegraphics[width=\linewidth]{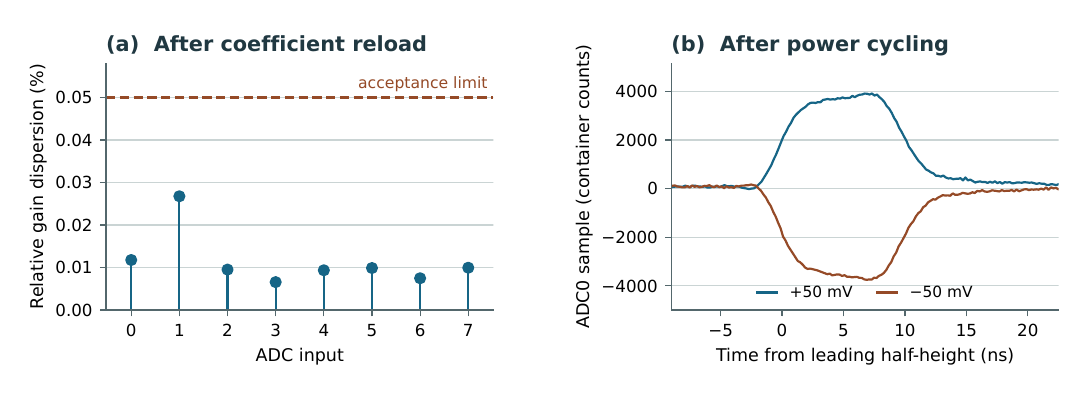}
\caption{Two independent checks of restoration. (a) Relative gain
dispersion from the commissioning sine measurement after separately
reloading each channel; the dashed line is the acceptance limit.
(b) First recorded input-0 trace of each polarity from the subsequent
supply-cycle test, with nominal 10~ns, 50~mV pulses applied after
start-up. Each trace is aligned to its own leading half-height crossing
for display. The ordinate is the stored sample value, without baseline
subtraction, filtering or averaging; the abscissa is not an absolute
trigger-delay measurement.}
\label{fig:cal-witness}
\end{figure}

Table~\ref{tab:cal-operation} distinguishes readback of the stored words
from verification of the analogue response. Figure~\ref{fig:cal-witness}(a)
shows the dispersion measured after reloading each channel;
figure~\ref{fig:cal-witness}(b) shows the input-0 pulse response after a
subsequent supply cycle. The mean full widths at half maximum were
10.05~ns for negative pulses and 10.06~ns for positive pulses (four
records of each polarity). The generator was connected but disabled during
power-up. This test establishes neither a cold-start success probability
nor performance with arbitrary live inputs or across the full operating
temperature range.

\begin{table}[htbp]
\centering
\caption{Checks at distinct stages of DC-carrier operation. Reading all
eight coefficient sets verifies the digital configuration; only a driven
input tests the resulting analogue waveform. These calibration records
contain 40\,960 samples per channel, rather than the 4096 used in
high-rate events.}
\label{tab:cal-operation}
\small
\renewcommand{\arraystretch}{1.20}
\begin{tabular}{@{}p{0.19\linewidth}p{0.34\linewidth}p{0.40\linewidth}@{}}
\hline
Stage & Test & Observation and scope\\
\hline
\raggedright Channel calibration & \raggedright Restore each input separately; repeat the same
sine measurement. & \raggedright All eight passed; relative gain dispersion
0.0066--0.0268\% after loading.\tabularnewline
\raggedright Powered hold and reset & \raggedright Hold input~0 for 608~s; separately reinitialize
the converter and reload its coefficients. & \raggedright The stored values were
retained during the hold and were recovered after reset without retraining.
This was not a temperature sweep.\tabularnewline
\raggedright Supply cycle & \raggedright Cycle the 12~V supply once with the source
connected but off. No host download or training tone. & \raggedright All eight sets were
restored automatically. Four full-table checks accompanied 32 eight-channel
records; only input~0 was driven.\tabularnewline
\raggedright Capture and reconnect & \raggedright Apply DC levels and pulses of both polarities;
disconnect and reconnect the acquisition client. & \raggedright After power-up,
$R_8$ was 8.63 and 10.14 counts at $+100$ and $-100$~mV, respectively. At zero after reconnection,
$R_8=10.31$ counts.\tabularnewline
\hline
\end{tabular}
\end{table}

\section{Analogue performance}
\label{sec:analogue}

\subsection{Input response and dynamic range}
The same ZYNQBeeRF7 module and the same 5~m RG-316 assembly were used
with the two carriers. Two measurements served different purposes. For a
frequency-response measurement the generator power was held constant; for
a dynamic-range measurement it was adjusted to obtain a specified
digitized carrier amplitude. Adjusting the power in the first measurement
would mask the attenuation being measured. Conversely, fixed source
power would excite the same converter to different fractions of full
scale on the two carriers.

For each spectral record, 40\,960 samples per channel were acquired
at 5~GSa/s. A periodic four-term Blackman--Harris window was applied.
Carrier power was integrated over the peak bin and five bins on either
side. The SNR calculation excludes DC and the folded second
through tenth harmonics; the signal-to-noise-and-distortion ratio (SINAD)
includes harmonic distortion. The spurious-free dynamic range (SFDR) is
reported relative to the carrier, not to full scale. For a carrier level
$L$ in dBFS we also report $\mathrm{SNR}_{\mathrm{FS}}=\mathrm{SNR}-L$,
a normalization that assumes constant noise power. It is not a measurement
at full scale. Long spectral records were used here to resolve small
tones; they are distinct from the 4096-sample records used in the
trigger-rate experiments. Table~\ref{tab:dynamic} gives the results
near $-1$~dBFS.

\begin{table}[hbp]
\centering
\caption{Eight-channel ranges at 100.317~MHz with the same source cable.
Each range is the minimum and maximum across channels, not a measurement
uncertainty. No extrema from different columns are assumed to belong to
the same channel.}
\label{tab:dynamic}
\small
\begin{tabular}{lcc}
\hline
Quantity & DC-coupled carrier & AC-coupled carrier\\
\hline
Digitized carrier (dBFS) & $-1.034$ to $-0.971$ & $-1.100$ to $-1.079$\\
Measured SNR (dB)         & 54.10--57.09 & 57.68--58.84\\
Full-scale-normalized SNR (dB) & 55.12--58.07 & 58.78--59.93\\
SINAD (dB)                & 54.05--57.02 & 57.50--58.57\\
SFDR (dBc)                & 61.36--70.17 & 69.02--72.49\\
\hline
\end{tabular}
\end{table}

\begin{figure}[tbp]
\centering
\includegraphics[width=\linewidth]{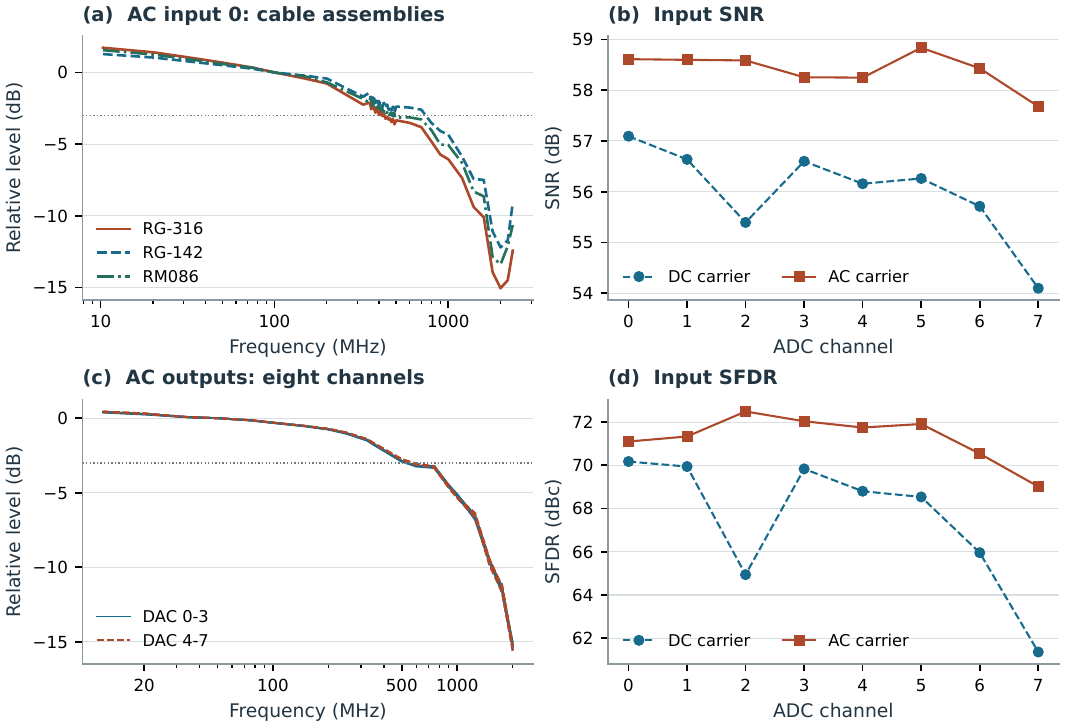}
\caption{(a) Response above 10~MHz of the same AC input with three 5~m
cable assemblies, each normalized to its own value at 100.317~MHz. The
absolute differences quoted in the text are at equal source power,
before normalization. (b) Measured SNR across the two carriers at
100.317~MHz near $-1$~dBFS. (c) AC output spectra: carrier levels
for eight DACs, relative to each output's nominal 50~MHz level.
(d) Input SFDR for the same records as in (b). Lines connecting
discrete frequency measurements guide the eye.}
\label{fig:analogue}
\end{figure}

The AC assembly gave the higher measured SNR on every channel at this
frequency. Figure~\ref{fig:analogue} compares the cable assemblies in
panel (a), the two carriers' input SNR and SFDR in panels (b) and (d),
and the AC carrier's DAC output response in panel (c).
The comparison is between two complete front ends: gain, filtering and
the calibration state were not identical. It does not isolate the effect
of a coupling capacitor or a transformer.
At 900.317~MHz the 5~m RG-142
and RM086 assemblies increased the received level by 2.30 and 0.97~dB,
respectively, relative to the 5~m RG-316 at the same generator setting.
The first two assemblies used the same short SMA--MCX adapter; the RG-316
assembly terminated directly in MCX. These are end-to-end differences,
not de-embedded cable losses. With the RG-316 and response normalized at
100.317~MHz, the first upper $-3$~dB crossing ranged from 417 to 433~MHz
on the AC carrier and from 594 to 609~MHz on the DC carrier. A crossing
relative to a specified frequency is not a bandwidth specification for
the bare converter, particularly for a response with a shallow minimum.

\subsection{Coupling between input channels and DAC output response}
To measure interchannel coupling on the AC carrier, one input at a time
was driven to approximately $-6$~dBFS and the other seven MCX inputs
were terminated locally in 50~$\Omega$. Each acquisition sampled the
driven and all seven undriven channels simultaneously. At the three
frequencies reported here, each estimate used 16 source-on frames and
24 source-off frames. A complex victim-to-driven tone ratio was formed
at the source frequency. The
detection threshold was the larger of two amplitude estimates: three
times the 95th percentile of the source-off victim tone referred to the
driven amplitude, and three standard errors of the complex ratio. For a
non-detection we report the empirical detection limit for that pair,
not a measured coupling value or a statistical confidence bound.

\begin{figure}[tbp]
\centering
\includegraphics[width=.98\linewidth]{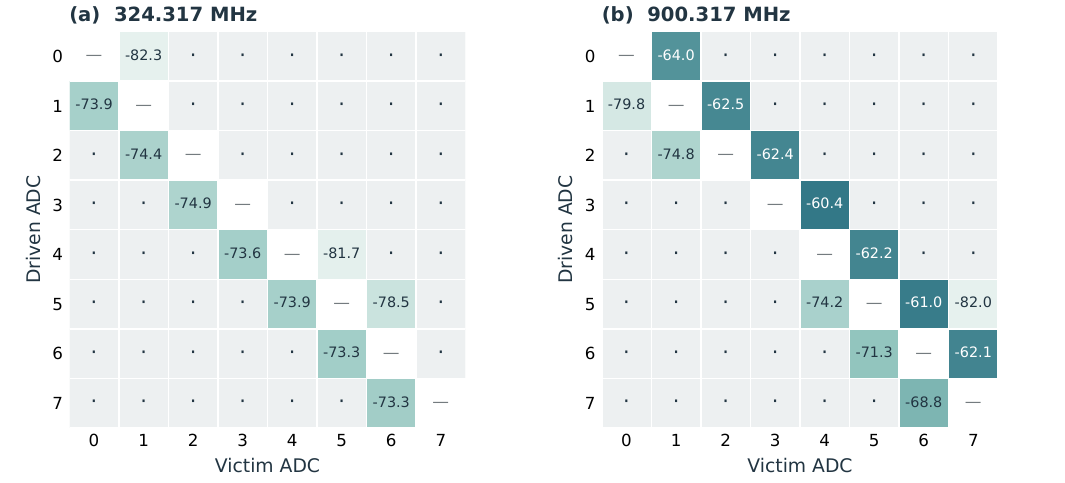}
\caption{Coupling on the AC carrier with all seven undriven inputs
terminated in 50~$\Omega$. Rows are driven inputs; columns are victims.
Numbers denote detected tones in dBc. A dot denotes a non-detection at
the limit for that pair; the diagonal is excluded. The reference for
dBc is the digitized carrier on the driven input in the same acquisition.}
\label{fig:xtalk}
\end{figure}

None of the 56 directed pairs was detected at 100.317~MHz; the least
restrictive upper bound was $-79.43$~dBc.
At 324.317~MHz ten pairs were detected, the strongest being input
7 into input 6 at $-73.28$~dBc (figure~\ref{fig:xtalk}(a)).
At 900.317~MHz input 3 into input 4 measured $-60.41$~dBc,
whereas the reverse direction was not detected at its $-82.34$~dBc
limit (figure~\ref{fig:xtalk}(b)).
In a separate reversible test at the same frequency, terminating only
the victim instead of leaving it open reduced coupling from input 1 to
2 by 4.75~dB and from 2 to 1 by 9.60~dB, with negligible change in
driven amplitude. The measurements describe directed signal transfer
between digitized channels under stated loads. They are not a measurement
of reciprocal two-port scattering parameters and do not identify a
particular connector, transformer or board trace as the cause.

For a separate DAC measurement, the eight AC outputs were connected in
turn to a 50~$\Omega$ spectrum analyser (Rohde \& Schwarz FSV30)
through the same 2~m RG-316 cable. Each channel replayed a
40\,960-sample sinusoid at each of 20 frequencies with peak amplitudes
of 1024 and 2048 counts in the signed sample container. The 320 carrier measurements
formed part of 592 records including controls. At 50.049~MHz and
2048 counts, the carrier power at the analyser corresponded to
34.44--35.66~mV peak-to-peak across 50~$\Omega$. Linear interpolation
between the measured points bracketing the first $-3$~dB crossing
gave 525--581~MHz relative to the nominal 50~MHz reference. No correction
for cable loss was applied. At these small amplitudes a wide-span trace
cannot establish an intrinsic DAC spur limit below the analyser floor.

\section{Event acceptance and sustained readout}
\label{sec:rates}
\subsection{Two triggers within one acquisition window}
\begin{figure}[tbp]
\centering
\includegraphics[width=.94\linewidth]{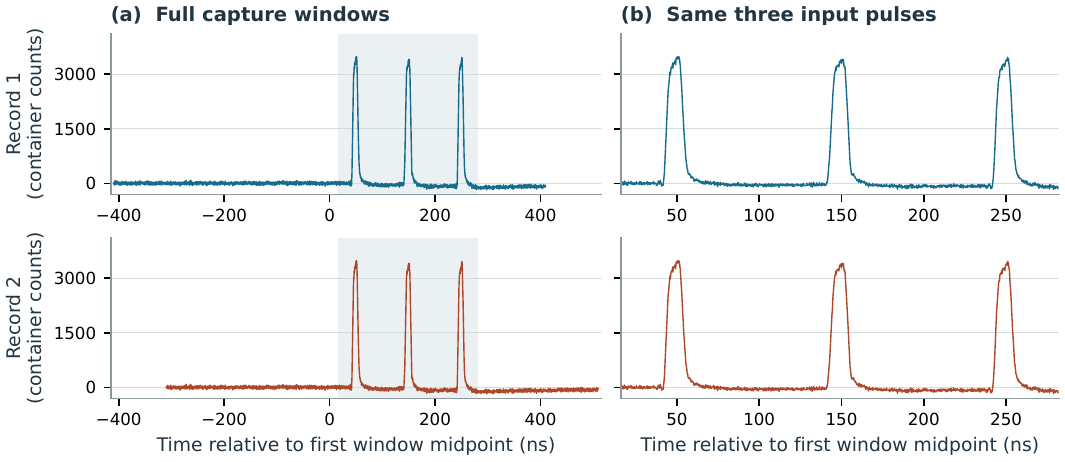}
\caption{Records from two accepted triggers in a three-pulse electrical
input, plotted from the stored ADC0 samples on a common time axis.
Zero is the midpoint of the first window, not a fitted pulse time.
The left panels show the full records; the right panels enlarge the
same interval. The second time series is translated using the recorded
time tags. No filtering or baseline subtraction was applied.}
\label{fig:neighbours}
\end{figure}

Short pulse trains and uninterrupted periodic excitation probe different
limits. With both capture banks initially available, each of four
two-pulse trains with a spacing of 100~ns produced two events. In the
three-pulse example in figure~\ref{fig:neighbours}(a), three threshold
crossings were counted, but only the first two were accepted. They
produced separate records with time tags 100~ns, or 500 sample intervals, apart.
After translation by this amount, the overlapping 3596 sample values
per channel are identical, including the baseline fluctuations, on all
eight channels. The third pulse is visible in both accepted waveforms
(figure~\ref{fig:neighbours}(b)).
Its presence there is not a third accepted trigger. Neither the 819.2~ns
window length nor the reciprocal of a sustained trigger rate is a fixed
two-pulse dead time.

\subsection{Tests with and without host pauses}
The source was a periodic electrical pulse train applied to ADC0 with
threshold triggering. For each run we compared the numbers of source
crossings, accepted triggers and events verified in host RAM. We also
checked event sequence, the cyclic redundancy check (CRC), receiver
discards and final queue occupancy. The host checked events during
acquisition; these rates are not specifications for archival disk
writes. A run with a busy rejection is not called loss-free merely
because all accepted events passed their CRC.

On the DC carrier, the receiver with event prefetch sustained 170~kHz
in three 30~min runs with a total of 917\,999\,364 verified events and
144 host pauses. A 60~s run at 175~kHz verified 10\,502\,680 events on
this carrier. The subsequent long runs in table~\ref{tab:runs} used the
AC carrier and the same receiver prefetch design. The Linux host had
a Ryzen 9 7945HX processor, 64~GB of RAM and a PCIe Gen4 $\times$8
interface. Four card-to-host readers and two checking threads used a
private, locked 17.1~GB host buffer; at least 95\% of the buffer was
backed by transparent huge pages before and after each run. The receiver firmware
and DMA driver were unchanged; device requests remained 4~KiB, irrespective
of the 2~MiB pages backing the host buffer. With matched initial backlogs
and a continuing 175~kHz input, the host read and acknowledged data at
12.580~GB/s with the original buffer and 13.202~GB/s with the new one.
The arriving records required 12.1856~GB/s: the headroom for clearing
the queue rose from 0.394 to 1.016~GB/s. The original host configuration
had already completed two uninterrupted 30~min runs at 179.5~kHz, with
peak queues of 27\,551 and 52\,211 slots; the new buffer peaked at 7716
at that rate. Lower occupancy, not just completion of a run, distinguishes
the two host configurations.

\begin{table}[tbp]
\centering
\caption{AC-carrier readout with periodic excitation. Rate is the nominal
generator setting. Rejected is the number of triggers refused at the source; peak
is the maximum number of occupied receiver event slots out of 61\,440.
All accepted events were verified; no receiver discard or CRC mismatch
was recorded in any row. The interrupted run and its repeat are both
listed.}
\label{tab:runs}
\small
\begin{tabular}{r l r r r r}
\hline
Rate (kHz) & Host condition & Time (s) & Verified events & Rejected & Peak\\
\hline
175 & Pauses only        & 1800 & 315\,001\,621 & 0 & 48\,004\\
175 & Pauses + CPU load  & 1800 & 315\,006\,985 & 0 & 48\,949\\
175 & Pauses + memory load & 958.05 & 167\,664\,132 & 3 & 47\,962\\
175 & Same, repeated     & 1800 & 314\,999\,258 & 0 & 48\,118\\
179.5 & Unperturbed      & 1800 & 323\,098\,783 & 0 & 7716\\
179.7 & Unperturbed      & 1800 & 323\,452\,631 & 0 & 4279\\
180 & Unperturbed        & 0.02036 & 3704 & 7 & 992\\
\hline
\end{tabular}
\end{table}

Each qualifying 175~kHz run included twelve host pauses at each of
10, 50, 100 and 250~ms. In the three qualifying runs, 945\,007\,864
events were verified, with no source rejection. Each queue drained at
the end. A 250~ms pause alone admits about 43\,750 additional events
at this rate, explaining why a high queue level is expected. The
slowest recovery after this pause length, to at most 1024 occupied
slots, was 4.18~s. In the first memory-load run three source rejections
were observed after about 958~s. A stored waveform contained an
unexpected transient approximately 0.82~ns wide at half height; its
origin was not established. The repeat passed under the original
criteria. We retain the interrupted run rather than attributing its
rejections to either the host or an identified external noise source.

Without imposed load or pauses, the highest tested loss-free 30~min
point was 179.7~kHz. At that nominal rate the samples account for
11.777~GB/s, the Aurora user port for 97.895~Gbit/s and the host
slots for 12.513~GB/s. These are event-size budgets, not measurements of
serial lane utilization. At 180~kHz the source counted 3711 crossings
and rejected seven. All 3704 accepted events were checked, and the
receiver queue used less than 2\% of its capacity in this short run.
The first discrepancy thus arose before the receiver and PCIe readout.

\subsection{Transmitter service at 180 kHz}
\begin{figure}[tbp]
\centering
\includegraphics[width=\linewidth]{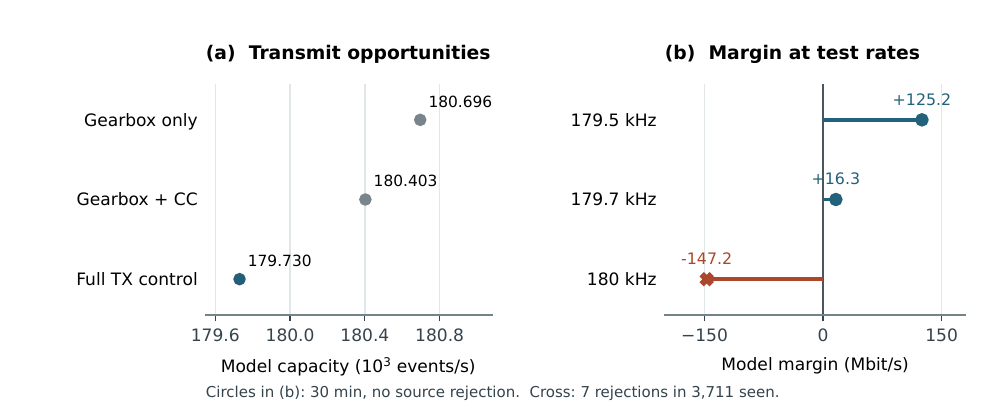}
\caption{(a) Modelled event capacity obtained by counting transmit
opportunities with the gearbox alone, with clock compensation (CC), and
with full transmitter control, including channel bonding. (b) Modelled
user-data capacity minus the required event traffic at each nominal
rate. Circles mark completed, rejection-free 30~min tests; the cross
marks a short test with seven source rejections. Neither panel reports
a measurement of physical-link bandwidth.}
\label{fig:budget}
\end{figure}

An event occupies 2128 words at the 256-bit Aurora transmit interface.
The interface does not accept a user word on every clock. We counted
the available cycles in the generated transmitter over 19 complete
channel-bonding intervals with downstream backpressure removed. Channel
bonding periodically inserts control information to align the link lanes;
these cycles are unavailable for user data.
Of 3\,210\,691 clocks, 3\,048\,361 were available for user data.
With $f_{\mathrm{tx}}=402.83203125$~MHz, the corresponding capacity is
\begin{equation}
 r_{\mathrm{tx}} =
 \frac{f_{\mathrm{tx}}}{2128}
 \frac{3\,048\,361}{3\,210\,691}
 =179.730~\mathrm{kHz}.
 \label{eq:capacity}
\end{equation}
Equation~\eqref{eq:capacity} gives the event capacity for the full
transmit schedule in figure~\ref{fig:budget}(a). Expressed as a user
data rate, it is 97.911~Gbit/s, whereas the event format requires
98.058~Gbit/s at 180~kHz (figure~\ref{fig:budget}(b)). Even this
small deficit eventually delays the return of a source capture bank.
It does not imply a lost event at every bonding interval.

In a cycle-resolved simulation retaining the generated transmit schedule,
1398 of 1400 triggers were accepted at 180~kHz, compared with 1400 at
179.7~kHz. In a counterfactual simulation at 180~kHz, suppressing only
the periodic channel-bonding gate on transmit readiness restored all
1400 events. This modification was not made to the hardware. In the
unaltered simulation both capture banks awaited service at rejection;
on hardware the busy counter and gaps at intervals of 528 or 529 input
triggers provided independent evidence, not a cycle-by-cycle observation
of the bank states. The two results localize the failure for this source,
firmware and frame format. They do not measure the peak mixed read/write
bandwidth of the receiver memory.

\section{Discussion and conclusions}
The measurements distinguish waveform fidelity from event acceptance.
On the DC front end the periodic error that grew with applied voltage
was strongly reduced by a calibration obtained with a sinusoid, without
flattening or subtracting the physical baseline. Once obtained, all eight
sets of coefficients were recovered after power cycling without applying
a training signal. That distinction matters when the available input is
a sparse pulse sequence rather than a continuous calibration tone.
The near-full-scale
spectra and the directional coupling matrix characterize what was
actually digitized at defined signal levels and terminations. They
are not measurements of a converter or a cable in isolation.

The main design result is that placing events in proportion to memory
width is insufficient if host page requests still determine when each
memory works. In the receiver, whole-burst buffering and then event-level
prefetch raised the sustained operating point to 150 and 170~kHz,
respectively. With the receiver fixed, a host allocation change provided
more bandwidth to clear a backlog \emph{while new events were arriving};
it enabled the 175~kHz pause tests. The 179.7~kHz uninterrupted result
answers a different question. At 180~kHz the first loss was upstream
of both the receiver and the host. Additional buffering could absorb a
finite burst but not a sustained deficit in transmitter capacity. The
bench measurements do not establish the acceptance for random arrivals,
absolute timing performance with a detector, or a sustained archival
write rate.

\section*{Data and software availability}
The underlying waveform files, acquisition logs, firmware and analysis
software have not been deposited in a public repository.

\acknowledgments
The author thanks Y.~Xue and J.~Zhang, technical engineers in the
electrical workshop of the Department of Engineering Physics at Tsinghua
University, for their skilled soldering and patient assistance with board
assembly, without which these measurements would not have been possible.

OpenAI Codex assisted with organizing the bench records, preparing plotting
and verification scripts, and drafting and editing the manuscript.

\bibliographystyle{JHEP}
\bibliography{references}
\end{document}